\documentclass[twocolumn, showpacs, amsmath, amssymb, prd]{revtex4}

\def\a{\alpha}
\def\b{\beta}
\def\g{\gamma}
\def\d{\delta}

\def\vt{\vartheta}
\def\L{{\cal {L}}}

\begin{document}

\title{Maxwell's field coupled nonminimally to quadratic torsion:
  Induced axion field and birefringence of the vacuum }

\author{Yakov Itin$^{1}$} 
\author{Friedrich W.\  Hehl$^{2}$}

\affiliation{{$^{1}$Institute of Mathematics, Hebrew University of
  Jerusalem and Jerusalem College of Engineering, Jerusalem 91904,
  Israel, email: {\tt itin@math.huji.ac.il}}}
\affiliation{$^{2}$Institute for Theoretical Physics, University of Cologne,
50923 K\"oln, Germany, and Department of Physics and Astronomy,
University of Missouri-Columbia, Columbia, MO 65211, USA, email: {\tt
  hehl@thp.uni-koeln.de}}

\begin{abstract}
  We consider a possible (parity conserving) interaction between the
  electromagnetic field $F$ and a torsion field $T^\alpha$ of
  spacetime. For generic elementary torsion, gauge invariant coupling
  terms of lowest order fall into two classes that are both nonminimal
  and {\it quadratic\/} in torsion.  These two classes are displayed
  explicitly.  The first class of the type $\sim F\,T^2$ yields
  (undesirable) modifications of the Maxwell equations. The second
  class of the type $\sim F^2\,T^2$ doesn't touch the Maxwell
  equations but rather modifies the constitutive tensor of spacetime.
  Such a modification can be completely described in the framework of
  metricfree electrodynamics. We recognize three physical effects
  generated by the torsion: (i) An axion field that induces an {\em
    optical activity\/} into spacetime, (ii) a modification of the
  light cone structure that yields {\em birefringence\/} of the
  vacuum, and (iii) a torsion dependence of the {\em velocity of
    light.\/} We study these effects in the background of a Friedmann
  universe with torsion.  {\it File tor17.tex, 02 August 2003}
\end{abstract}


\maketitle

\newpage

\section{Maxwell's field coupled to torsion}

Since torsion \cite{conv1} $T^a=(1/2)T_{bc} {}^a\vt^b \wedge \vt^c$ had been
introduced into gauge field theories of gravitation as a natural
ingredient, its possible interaction with the electromagnetic field
$F=(1/2)F_{ab}\vt^a \wedge \vt^b$ was discussed \cite{RMP}. 
More recent investigations include
\cite{Benn,Hammond,Puntigam,IAS,Mukho,Preuss,ROH}. Such a coupling,
provided we subscribe to Maxwellian electrodynamics including charge
and flux conservation \cite{Puntigam}, has to respect the {\em
  diffeomorphism\/} invariance of the underline geometry as well as
the {\em gauge\/} invariance of electrodynamics.

In the framework of the Einstein-Cartan \cite{RMP} and the Poincar\'e
gauge theory of gravity \cite{gron96}, there doesn't exist a
``comma-goes-to-semicolon'' rule $(\;)_{,a}\longrightarrow(\;)_{;a}$
or $\partial_a\longrightarrow\nabla_a$ \cite{MTW}, since this
would violate gauge invariance:
\begin{equation}\label {rule1}
  F_{ab}=2\partial_{[a}A_{b]}\quad \longrightarrow \quad
  2\nabla_{[a}A_{b]}=F_{ab} -T_{ab}{}^cA_c\,.
\end{equation}
The new field strength $2\nabla_{[a}A_{b]}$ is modified by a gauge
dependent term $-T_{ab}{}^cA_c$, and the Lagrangian constructed by
means of $\partial_a\longrightarrow\nabla_a$ would read, in
symbolic form,
 \begin{equation}\label {min4}
L\sim F_{ab}F^{ab} +2F\cdot A\cdot T+A\cdot A\cdot T\cdot T\,.
\end{equation}
Clearly, this $L$ is defective because of its lack of gauge invariance
under $A_a\longrightarrow A_a+\partial_a\phi$.

As a consequence of these considerations, the {\em free\/} Maxwell
Lagrangian is added to the gravitational Lagrangian with the unchanged
field strength $F_{ab}=2\partial_{[a}A_{b]}$. Nevertheless, in an
exact solution of the gravity-Maxwell system, the torsion may depend
on the electrical charge, as is exemplified by the
Reissner-Nordstr\"om solution with torsion in the framework of a
Poincar\'e gauge theory \cite{Reissner}. In other words, torsion is
influenced by the electric charge {\em in\/}directly via the
frame, the metric, and the field equations. This is what we dub {\em
  minimal coupling\/} in this exceptional case. What in the Lagrangian
looks like ``no interaction at all,'' still yields an effective
``minimal'' interaction.

In this article, we will describe a complete family of Maxwell-torsion
Lagrangains that are diffeomorphism and gauge invariant. We restrict
ourselves to {\em parity preserving\/} Lagrangians involving the
electromagnetic field strength and the torsion up to biquadratic
expressions. Parity violating Lagrangians \cite{Mukho} can be
studied in a similar manner. Thus we will display the family of parity
preserving Maxwell-torsion Lagrangians that are {\em non-minimally
  coupled\/} to lowest order \cite{Teys}.

In order to describe the complete family of possible coupling terms,
we introduce a pseudo-orthonormal coframe $\vt^a$ and express the
variables relative to this cobasis.  The electromagnetic potential
1-form reads $A=A_a\vt^a$, the electromagnetic field strength 2-form
$F=dA=(1/2)F_{ab}\vt^a\wedge\vt^b$, and the torsion 2-form
$T^a=(1/2){T_{bc}}^a\, \vt^b\wedge\vt^c$.  We represent the Lagrangian
expressions in symbolic form, for instance, $L\sim A\cdot F\cdot T$,
where the "$\cdot$"-sign denotes the contraction of the indices by
mean of the Lorentz metric $g_{ab}={\rm diag}(1,-1,-1,-1)$. We find the
following terms:\medskip

\noindent (i) Expressions of the type $A\cdot T$. The Lagrangian is 
only diffeomorphism invariant and reads explicitly $A_aT^{ab}\,_{b}$.
Such an expression is gauge invariant if and only if
$T^{ab}\,_{b,a}=0$. This condition is not fulfilled in general.\medskip

\noindent (ii) Expressions of the type $A\cdot T\cdot T$ and 
$F\cdot T$ involve an odd numbers of free indices. By using $g_{ab}$,
they cannot be contracted to a scalar. Consequently a diffeomorphism
invariant Lagrangian of such a type does not exist.\medskip

\noindent (iii) Expressions of the type $F\cdot T\cdot T$ have 8 
free indices. Hence they yield a family of diffeomorphism and gauge
invariant Lagrangians.  Taking into account the antisymmetries and
considering the possible combinations of indices, we obtain a linear
combination of elementary Lagrangians
\begin{eqnarray}\label{fttgen}
  L&=&\frac 12 F_{ab}\big(\alpha_1 T^{ab}\,_cT^{cm}\,_m+
  \alpha_2T^{anb}T_{nm}\,^m \nonumber\\ &&\qquad +\alpha_3 T
  ^{amn}T^b\,_{nm}+ \alpha_4 T^{amn}T_{mn}\,^b\big)\,.
\end{eqnarray}

\noindent (iv) Expressions of the types $A\cdot A\cdot T$ and 
$F\cdot F\cdot T$ have an odd number of free indices, so they do not
yield diffeomorphism invariant Lagrangians.\medskip

\noindent (v) Expressions of the type $A\cdot F\cdot T$ have 6 free 
indices, so they may serve as building blocks for diffeomorphism
invariant Lagrangians.  Considering the possible combinations of
indices, we obtain the following generic expression
\begin{equation}\label {aft1}
  L=F_{ab}\big(\beta_1A^aT^{bm}\,_m+\beta_2A_mT^{abm}
  +\beta_3A_mT^{amb}\big)\,.
\end{equation}
For imposing gauge invariance on this Lagrangian, it is necessary to
require some constraints for the torsion field as, for instance,
\begin{equation}\label {aft2}
\beta_2T^{abm}+\beta_3T^{m[ab]}=0\,.
\end{equation}
However, this would be an ad hoc procedure.\medskip

\noindent (vi) An expression of the type $A\cdot A\cdot T\cdot T$ 
yields a rich family of diffeomorphism invariant Lagrangians.
However, because of lack of gauge invariance, it is necessary to apply
a constraint of the form $([A\cdot T\cdot T]_a)^{,a}$.

\noindent If we look back to the Lagrangian (\ref{min4}), we recognize that
(\ref{min4}) is of type (v) and (vi) and thus has to be excluded.\medskip

\noindent (vii) The expressions of the type $A\cdot F\cdot T\cdot T$ 
have an odd number of free indices and, thus, do not yield
diffeomorphism invariant Lagrangians.\medskip

\noindent (viii) The expression of the type $F\cdot F\cdot T\cdot T$ has 
an even number of free indices and thus yields a family of
diffeomorphism and gauge invariant Lagrangians. We will construct the
complete list of such Lagrangians in the following sections.\medskip

Our considerations point to the existence of three different types of
Maxwell-torsion Lagrangians:\medskip

\noindent$\bullet$ Diffeomorphism invariant Lagrangians which do not
respect gauge invariance, namely the cases (i), (v), and (vi). Observe
that even in these cases the situation is not completely hopeless. The
gauge invariance could be reestablished by considering a torsion field
that fulfills a suitable constraint. For generic elementary torsion
such a constraint is unnatural and cannot be justified. However, if
the torsion is not an elementary field but rather constructed from
some other fields (scalar, vector, spinor, Kalb-Ramond, ... ), then a
gauge invariant Lagrangian may exist. An interesting possibility of a
torsion generated by an antisymmetric Kalb-Ramond field was studied 
recently \cite{Mukho}.\medskip

\noindent$\bullet$ The family (iii) of diffeomorphism and gauge
invariant Lagrangians.  On the level of the field equation, these
models yield an addition term that is independent of the
electromagnetic field.  Such a term can be treated as an additional
current.  Consequently, the corresponding model admits the existence
of a global (cosmological) electromagnetic field even without charges.
Such a modification of classical electrodynamics seems to be
unwarranted.\medskip

\noindent$\bullet$ The family of models with a supplementary
Lagrangians of type (viii). Eventually, this candidate is left over and
will be studied subsequently.

\section{Quadratic torsion coupled to Maxwell's field}

Recently Preuss \cite{Preuss}, in the context of a cosmological test
of Einstein's equivalence principle, suggested to investigate a
specific nonminimally coupled Lagrangian density $\L=L *1$ of the form
\begin{equation}\label{eq4}
  {}^{\rm(Pr)}\L=\ell^2*(T_a\wedge F)\, T^a\wedge F\,,
\end{equation}
which is supposed to be added to the conventional Maxwell Lagrangian.
Here $\ell$ is a coupling constant with the dimension of length and
the star $*$ denotes the Hodge duality operator. 

The Preuss Lagrangian is a specific example of the family (viii) of
the last section. In symbolic form, the additional torsion dependent
Lagrangian of type (viii) reads
\begin{equation}\label{eq5}
  {}^{\rm(tor)}L=-\frac 18 \ell^2\mathop{\sum}_{a,\cdots
    ,q}\Big(F_{ab}F_{cd}\,T_{klm}T_{npq}\Big)\,,
\end{equation}
where the summation is performed by contracting the indices with the
help of the metric tensor. The indices of the expression
$F_{ab}F_{cd}\,T_{klm}T_{npq}$ can be contracted in the following
three ways:\medskip

\noindent (i) All indices of the $F$-pair and of the $T$-pair are 
contracted separately. The only possibility for the $F$-pair is
$F_{ab}F^{ab}$.  For the $T$-pair we have three possibilities and,
accordingly, three independent Lagrangians:
\begin{eqnarray}
\label{lag1}
{}^{(1)}L&=&F_{ab}F^{ab}\, T_{mnk}T^{mnk}\,,\\
\label{lag2}
{}^{(2)}L&=&F_{ab}F^{ab}\,T_{mnk}T^{nkm}\,,\\
\label{lag3}
{}^{(3)}L&=&F_{ab}F^{ab}\,{T_{mn}}^n{T^{mk}}_k\,.
\end{eqnarray}
Certainly, these three Lagrangians could be rewritten via the
irreducible pieces of the torsion \cite{hehl95} or as the independent
parts of the teleparallel Lagrangian \cite{itin}.\medskip

\noindent (ii) Two free indices of the $F$-pair are contracted with 
two free indices of the $T$-pair. The only possible $F$-pair is
$F_{am}{F_b}^m=F_{bm}{F_a}^m$. Thus the $T$-pairs have to be symmetric
too. Consequently, we have five independent Lagrangians:
\begin{eqnarray}
\label{lag4}
{}^{(4)}L&=&F_{am}{F_b}^m\,T^{anb}{T_{nk}}^k \,,\\
\label{lag5}
{}^{(5)}L&=&F_{am}{F_b}^m\,T^{akn}{T^b}_{kn}\,,\\
\label{lag6}
{}^{(6)}L&=&F_{am}{F_b}^m \,T^{akn}{T^b}_{nk} \,,\\
\label{lag7}
{}^{(7)}L&=&F_{am}{F_b}^m \,T^{akn}{T_{kn}}^b \,,\\
\label{lag8}
 {}^{(8)}L&=&F_{am}{F_b}^m\,{T^{ak}}_k{T^{bn}}_n \,.
\end{eqnarray}

\noindent (iii) The $F$-pair and the $T$-pair have four free indices. 
Taking the $F$-pair in the general form $F_{ab}F_{cd}$, we find the
following possibilities:
\begin{eqnarray}
 \label{lag9}
 {}^{(9)}L&=&F_{ab}F_{cd}\,{T^{am}}_mT^{bcd} \,,\\ 
 \label{lag10}
 {}^{(10)}L&=&F_{ab}F_{cd}\,{T^{am}}_mT^{cdb} \,,\\
 \label{lag11}
 {}^{(11)}L&=&F_{ab}F_{cd}\,{T_m}^{ab}T^{mcd} \,,\\
 \label{lag12}
 {}^{(12)}L&=&F_{ab}F_{cd}\,{T_m}^{ac}T^{mdb} \,,\\
 \label{lag13}
 {}^{(13)}L&=&F_{ab}F_{cd}\,{T_m}^{ad}T^{mbc} \,,\\
 \label{lag14}
 {}^{(14)}L&=&F_{ab}F_{cd}\,T^{mab}{T^{cd}}_m \,,\\
 \label{lag15}
 {}^{(15)}L&=&F_{ab}F_{cd}\,T^{mac}{T^{db}}_m \,,\\
 \label{lag16}
 {}^{(16)}L&=&F_{ab}F_{cd}\,T^{abm}{T^{cd}}_m \,,\\
 \label{lag17}
 {}^{(17)}L&=&F_{ab}F_{cd}\,T^{acm}{T^{db}}_m \,.
\end{eqnarray}

Summing up, the explicit form of the general torsion Lagrangian
(\ref{eq5}) reads
\begin{equation}\label{gentor}
  \frac {1}{\ell^2}\,\,^{\rm(tor)}L=-\frac 18\mathop{\sum}_{i=1}^{17}
  \gamma_{i}\,^{(i)}L\,,
\end{equation}
with the dimensionless constants $\gamma_i$.

Let us rewrite the Preuss Lagrangian (\ref{eq4}) in terms of
components.  With the formula $*w\wedge\vt_a=-*(e_a\rfloor w)$ (see
\cite{hehl95}), we find:
\begin{eqnarray}
  \frac {1}{\ell^2}\,\,^{\rm (Pr)}{\cal L}
&=&\frac
1{16}T_{bca}F_{mn}{T^{pq}}^aF^{rs}*\vt^{bcmn}\wedge\vt_{pqrs}\nonumber\\ 
&=&-\frac 1{16}T_{bca}F_{mn}{T^{pq}}^aF^{rs}*(e_p\rfloor
\vt^{bcmn})\wedge\vt_{qrs} \nonumber\\ &=&-\frac
1{8}T_{bca}F_{mn}{T^{pq}}^aF^{rs}*(\d^b_p\vt^{cmn}+\nonumber\\ 
&&\qquad \d^m_p\vt^{bcn})\wedge\vt_{qrs}=\cdots\nonumber\\ 
&=&\frac 14(T_{bca}T^{bca}F_{mn}F^{mn}+
4T_{bca}T^{bma}F_{mn}F^{nc}\nonumber\\
&&\qquad + T_{bca}T^{mna}F_{mn}F^{bc})*1\,.
\end{eqnarray}
Thus, the Preuss Lagrangian is given as a linear combination
according to
\begin{equation}\label{eq10}
  ^{\rm (Pr)}L=\frac{\ell^2}4\,
  \left(\,{}^{(1)}L-4\,{}^{(5)}L+\,{}^{(16)}L \right)\,,
\end{equation}
that is, $\gamma_1=-2\,,\,\gamma_5=8$, and $\gamma_{16}=-2$, all other
$\gamma_i$'s vanish.

\section{Electrodynamics with a local and linear ``constitutive law''}

Now we want to put our findings on the nonminimal coupling of torsion
to Maxwell's field into an efficient framework. This coupling can be
reformulated as standard Maxwell theory with an effective constitutive
tensor $\chi^{abcd}$ that is quadratic in the torsion.  Recently it
was shown \cite{IAS,gentle,drguillermo,book} that non-standard
elementary fields, namely axion, skewon, and dilaton fields can appear
even within the framework of linear electrodynamics.  The metricfree
Maxwell equations $dH=J$ and $dF=0$, with the excitation $H=({\cal H},
{\cal D})$ and the field strength $F=(E,B)$, remain untouched by this
construction. The same is true for the nonminimally coupled torsion
field. The modification appears in the so-called spacetime relation,
i.e., the relation between the 2-forms $H$ and $F$.  We write these
2-forms in components as
\begin{equation}\label{eq1}
F= \frac 12 F_{ab}\vt^{ab}\,, \qquad H=\frac 12 H^{ab}\eta_{ab}\,.
\end{equation}
The coframe $\vt^a$ is assumed here to be orthonormal relative to some
metric $g$. Moreover, $\vt^{ab}:=\vt^a\wedge\vt^b$ is a basis of the
space of untwisted 2-forms.  Correspondingly, if $e_a$ is the
orthonormal frame dual to the coframe $\vt^a$, then
$\eta_{ab}:=e_a\rfloor e_b\rfloor \rm{vol}$ is a basis of twisted 2-forms.
The linear homogeneous spacetime relation between the 2-forms $H$ and
$F$,
\begin{equation}\label{eq2}
H^{ab}=\frac 12 \chi^{abcd}F_{cd}\,,
\end{equation}
is characterized by the constitutive tensor $\chi^{abcd}$ with 36
independent components \cite{conv}.  This tensor can be decomposed
irreducibly according to
\begin{equation}\label{eq3}
  \chi^{abcd}=\widetilde{\chi}^{abcd}+\hat{\chi}^{abcd}+
  \check{\chi}^{abcd}\,,
\end{equation}
where $\widetilde{\chi}^{abcd}$ is the principal part (20 independent
components), $\hat{\chi}^{abcd}=(\chi^{abcd}-\chi^{cdab})/2$ is the
skewon part (15 independent components), and
$\check{\chi}^{abcd}=\chi^{[abcd]}=\theta\epsilon^{abcd}$ is the axion
part (1 component).  The principal part is known to be related to the
metrical structure of spacetime.  Moreover, as it is shown in
\cite{book}, the metric can even be derived from the tensor
$\widetilde{\chi}^{abcd}$. So, it is natural to expect that the
additional parts of the constitutive tensor may also be related to the
geometry of spacetime.

In fact, starting from the standard Lagrangian
\begin{equation}\label{eq6}
  L=-\frac 12H\wedge F=-\frac 14H^{ab}F_{ab}*1=-\frac
  18\chi^{abcd}F_{ab}F_{cd}*1\,,
\end{equation}
the additional constitutive tensor, induced by the
torsion, turns out to be
\begin{equation}\label{eq7}
  {}^{(\rm{tor})}\chi^{abcd}=\ell^2\Big[\sum_{k\cdots
    q}\Big(T_{klm}T_{npq}\Big)\Big]^{[ab][cd]}\,,
\end{equation}
where the summation is understood as a contraction of the indices by
means of the metric tensor such that the indices $a,b,c,d$ remain
free.

The expression (\ref{eq7}) is derived from a Lagrangian. Hence, in
addition to the antisymmetry relations
${}^{(\rm{tor})}\chi^{abcd}=-{}^{(\rm{tor})}
\chi^{bacd}=-{}^{(\rm{tor})}\chi^{abdc}$, it has to satisfy
${}^{(\rm{tor})}\chi^{abcd}={}^{(\rm{tor})}\chi^{cdab}$.  Accordingly,
a skewon piece does not occur in (\ref{eq7}) and the torsion field,
in the framework of this Lagrangian approach, will modify the
principal part and, additionally, generate an axion part of the
constitutive tensor.

The axion part takes the form
\begin{equation}\label{eq8}
  {}^{(\rm{tor})}\check{\chi}^{abcd}=\ell^2\Big[\sum_{k.\cdots
    q}\Big(T_{klm}T_{npq}\Big)\Big]^{[abcd]}\,.
\end{equation}
The modification of the principal part
\begin{equation}\label{eq8x}
  {}^{(\rm{tor})}\widetilde{\chi}^{abcd}={\ell^2}\Big[\sum_{k\cdots
    q}\Big(T_{klm}T_{npq}\Big)\Big]^{[ab][cd]}-{}^{(\rm{tor})}
  \check{\chi}^{abcd}
\end{equation}
may yield {\em birefringence\/} of the vacuum \cite{Preuss,ROH}. Let 
\begin{equation}\label{vactor}
\chi^{abcd}={}^{(\rm{vac})}\chi^{abcd}+ {}^{(\rm{tor})}\chi^{abcd}
\end{equation}
be prescribed, here $^{(\rm{vac})}\chi^{abcd}$ denotes the vacuum
constitutive tensor. Then the light propagation can be determined by
means of the generalized Fresnel equation
\begin{equation}\label{eq7x}
{\mathcal{G}}^{abcd}q_aq_bq_cq_d=0\,,
\end{equation}
where $q_a$ is the wave covector of a propagating surface of
discontinuity and
\begin{equation}\label{eq7xx}
{\mathcal{G}}^{abcd}:=\frac 1{4!}\epsilon_{mnpq}\epsilon_{rstu}
\chi^{mnr(a}\chi^{b|ps|c}\chi^{d)qtu}\,
\end{equation}
is the Tamm-Rubilar tensor density \cite{book}. Thus additional terms
in the constitutive tensor yield, via (\ref{eq7xx}), a modification of
the original Fresnel equation. In some cases, the original vacuum
light cone is split into {\em two\/} light cones yielding
birefrigence, as has been shown recently for the Preuss Lagrangian
\cite{Preuss,ROH}.

\section{Axion field induced by torsion}

The axion, that is, a pseudo-scalar field, was extensively studied in
different contexts of field theory. Its quantized version is believed
to provide a solution to the strong CP problem of QCD
\cite{Peccei:1977hh}. The emergence of axions is a general phenomenon
in superstring theory \cite{Green:sp}. For the role of the axion in
inflationary models, see e.g., \cite{Kallosh:1995hi}. The axion as a
classical field also appears in various discussions of the equivalence
principle in gravitational physics \cite{Dicke,Ni,Lammerzahl}. In the
context discussed in this paper, the axion field is induced by the
nonminimal coupling of Maxwell's field to the torsion of spacetime
according to (\ref{eq8}). Explicitly, we find the following:

\noindent (i) The Lagrangians (\ref{lag1})--(\ref{lag3}) yield 
\begin{equation}\label{eq8xxx}
   {}^{(\rm{tor})}\check{\chi}^{abcd}= \ell^2g^{[ac}g^{bd]}(T\cdot T)=0\,.
\end{equation}

\noindent (ii) The Lagrangians (\ref{lag4})--(\ref{lag8}) yield
\begin{equation}\label{eq8xx}
 {}^{(\rm{tor})}\check{\chi}^{abcd}= \ell^2 g^{[bd}(T\cdot T)^{ac]}=0\,.
\end{equation}

\noindent (iii) As for the Lagrangians from the third group 
(\ref{lag9})--(\ref{lag17}), we find the general form of the axion
field $\theta\!=\!\epsilon_{abcd}\, {}^{(\rm{tor})}\check{\chi}
^{abcd} /4!$ as
\begin{eqnarray}\label{eq9}
  \theta&=&-\ell^2\epsilon_{abcd}\Big(\a_1{T_m}^{ab}{T^{cdm}}+
  \a_2{T}^{mab}{T_m}^{cd}\nonumber\\ && \qquad\quad+
  \a_3{T^{ab}}_m{T^{cdm}}+\a_4T^{abc}{T^{dm}}_m\Big)\,,
\end{eqnarray}
where $\a_1,\cdots,\a_4$ are free dimensionless parameters, which are 
linear combinations of $\g$'s. 

The axion field of the Preuss Lagrangian reads explicitly
\begin{equation}\label{eq11}
{}^{({\rm Pr})}\check{\chi}^{abcd}=2\ell^2\,{T^{[ab}}_m{T^{cd]m}}\,
\end{equation}
or 
\begin{equation}\label{xxx}
  ^{\rm (Pr)}\theta=\frac{\ell^2}{12}\,
  \epsilon_{abcd}\,T^{abm}\,T^{cd}{}_m\,.
\end{equation}

If in vacuum an axion field $\theta$ emerges, then the Lagrangian
picks up an additional piece $\sim\theta\,F\wedge F$, see \cite{book}.
Accordingly, the inhomogeneous Maxwell equation reads
\begin{equation}\label{inhom}
  d\,^*F+d\theta\wedge F=J\,.
\end{equation}
Obviously, only a nonconstant axion field contributes. The homogeneous
Maxwell equation remains untouched: $dF=0$. Hence charge conservation
is guaranteed, that is, $dJ=0$.

The axion field doesn't influence the light cone structure, see
\cite{drguillermo,book}. However, as was shown by Haugan and
L\"ammerzahl \cite{clausannalen}, see also the literature given there,
the coupling of Maxwell's field to a nonconstant axion field, in the
case of a plane electromagnetic wave, amounts to a rotation of the
polarization vector of the wave, i.e., the axion field induces an {\it
  optical activity\/} (similar to a solution of sugar in water
\cite{Faraday}).

\section{Principal part of the constitutive tensor 
$\widetilde{\chi}^{abcd}$ and torsion}

A modification of the principal part of the constitutive tensor may
yield crucial changes in the structure of the light cone, as we
discussed at the end of Sec.III. In particular Preuss studied recently
possible observational consequences \cite{Preuss}. In the framework of
our model, the constitutive tensor is determined by the torsion of
spacetime, inter alia. Consequently, if the appropriate changes in the
propagation of light were observed, this would represent evidence for
the existence of the torsion field.

For different elementary Lagrangians we find the following effects on
the light cone structure (we put $\ell=1$ for simplicity):\medskip

\noindent (i) The Lagrangians (\ref{lag1})--(\ref{lag3}) yield 
\begin{equation}\label{eq10a}
 {}^{(\rm{tor})}  \widetilde{\chi}^{abcd}=S(g^{ac}g^{bd}-g^{ad}g^{bc})=
  2Sg^{a[c}g^{d]b}\,,
\end{equation}
where $S$ is a scalar function quadratic in torsion. Therefore, the
Tamm-Rubilar tensor density changes only by a conformal factor.
Accordingly, the structure of the light cone is preserved for such
models.\medskip

\noindent (ii) For the Lagrangians (\ref{lag4})--(\ref{lag8}), we 
introduce the abbreviation $S^{ab}=S^{ba}:=\big[T\cdot T\big]^{(ab)}$.
In this group of models, the axion field is absent. Thus
\begin{eqnarray}\label{eq10b}
 {}^{(\rm{tor})}  \widetilde{\chi}^{abcd}&=&\frac
  14\left(g^{ac}S^{bd}-g^{ad}S^{bc}+g^{bd}S^{ac}-g^{bc}S^{ad}\right)
  \nonumber\\ &=& g^{[a|[c}\,S^{d]|b]}\,.
\end{eqnarray}
In $6\times 6$ matrix form, the $3\times 3$ constitutive matrices, in
an orthogonal frame, read \cite{greek}:
\begin{eqnarray}\label{eq10c}
 {}^{(\rm{tor})}  A^{\a\b}&:=& {}^{(\rm{tor})}\chi^{0\b0\a}=\frac 14
  \left(g^{00}S^{\a\b}+g^{\a\b}S^{00}\right)\,,\nonumber\\ 
 {}^{(\rm{tor})}  B_{\a\b}&:=&\frac 14\epsilon_{\b\g\d}\epsilon_{\a\mu\nu}
 \, {}^{(\rm{tor})}\chi^{\g\d\mu\nu}=\frac
  14\left(g_{\a\b}S_\g{}^\g-S_{\a\b}\right)\,,\nonumber\\ 
 {}^{(\rm{tor})}  {C^\a}_\b&:=&\frac 12\epsilon_{\b\g\d}\, 
{}^{(\rm{tor})}\chi^{\g\d
    0\a}=\frac 14 {\epsilon^\a{}_{\b\g}}\, S^{0\g}\,,\nonumber\\ 
 {}^{(\rm{tor})}  {D_\a}^\b&:=&\frac 12 \epsilon_{\a\g\d}\, {}^{(\rm{tor})}
  \chi^{0\b\g\d} =\frac 14\epsilon^\b{}_{\a\g}\, S^{0\g}\,.
\end{eqnarray}
These matrices obey
\begin{equation}\label{eq10d}
  {}^{(\rm{tor})}A= {}^{(\rm{tor})}A^{\rm T}\,,\; {}^{(\rm{tor})}B=
  {}^{(\rm{tor})}B^{\rm T}\,,\; {}^{(\rm{tor})}C=
  {}^{(\rm{tor})}D^{\rm T}\,,
\end{equation}where ${\rm T}$ denotes the transposed of a matrix.
As shown in \cite{book}, these relations, together with the closure
condition for the constitutive tensor, guarantee uniqueness of the
light cone. Thus no birefringence emerges also in this group of
models.\medskip

\noindent (iii) For the Lagrangians from the third group 
(\ref{lag9})--(\ref{lag17}), birefringence is a generic property. An
example of this effect, for Preuss Lagrangian in the case of
spherically symmetric torsion, was given in \cite{ROH}.

\section{Example: Spatially homogeneous torsion in a Friedmann 
  cosmos}

As an example, we consider a spatially homogeneous torsion field
appearing in a Friedmann type solution of gauge theories of gravity
\cite{nara,goenner,albert}.  The independent nonvanishing components
of the torsion are
\begin{equation}\label{eq12}
  T_{0\a\b}=\frac u{\ell}\,\delta_{\a\b}\,,\qquad T_{\a\b\g}=\frac
  v{\ell}\,\epsilon_{\a\b\g}\,.
\end{equation}
where $u=u(t)$ and $v=v(t)$ are functions of time the explicit forms
of which depend on the specific cosmological model under consideration
\cite{nara,goenner,albert}.  We chose an orthonormal frame with metric
$g_{ab}={\rm diag}(+1,-1,-1,-1)$.

The general Lagrangian (\ref{gentor}), generated by torsion, yields,
together with the constitutive tensor of the vacuum, the following
nonvanishing components of the total constitutive tensor:
\begin{eqnarray}\label{eq13a}
&&\chi^{0\a\b\g}=(c_1uv)\,\epsilon^{\a\b\g}\,,\\
\label{eq13b}
&&\chi^{0\a0\b}=-(1+c_2u^2+c_3v^2)\,\d^{\a\b}\,,\\
\label{eq13c}
&&\chi^{\a\b\g\d}=(1+c_4u^2+c_5v^2)(\d^{\a\g}\d^{\b\d}-\d^{\a\d}\d^{\b\g})\,.
\end{eqnarray}
Here $c_1,\cdots, c_5$ are linear combinations of $\g$'s.

Accordingly, the torsion induces an axion field
$\theta=c_1 \, uv\,.$ 
In general, the time derivative $\dot{\theta}$ of this axion field
doesn't vanish. Thus in (\ref{inhom}) only an additional {\it time\/}
derivative shows up and the optical activity induced by the torsion of
the Friedmann cosmos is proportional to $ \dot\theta=c_1 \,
(\dot{u}v+u\dot{v})\,.$ The detailed form depends on the specific
Friedmann model under consideration.

A further physical effect originates from the principal part
$\widetilde{\chi}^{abcd}$ of the constitutive tensor.  For
(\ref{eq13a})-(\ref{eq13c}), the $3\times 3$ constitutive matrices
read
\begin{equation}\label{eq19x}
  A^{\a\b}=-f\d^{\a\b}\,,\; {D_\a}^\b={C^\b}_\a=g\d_\a^\b\,, \;
  B_{\a\b}=h\d_{\a\b}\,,
\end{equation}
where $f=1+c_2u^2+c_3v^2$, $g=c_1uv$, and $h=1+c_4u^2+c_5v^2$.
In a tedious calculation, the Tamm-Rubilar tensor
density ${\mathcal{G}}^{abcd}$ can be determined. With 
\begin{equation}\label{TR3D}
  M:={\mathcal{G}}^{0000}\,,\;M^{\a\b}:= 6{\mathcal{G}}^{00\a\b}\,,
  \;M^{\a\b\g\d}:={\mathcal{G}}^{\a\b\g\d}\,,
\end{equation}
we find the nonvanishing coefficients
\begin{equation}\label{eq19xx}
  M=-f^3\,,\quad\!\!  M^{\a\b}=2f^2h\d^{\a\b}\,,\quad\!\!
  M^{\a\b\g\d}=-fh^2\d^{(\a\b}\d^{\g\d)}\,.
\end{equation}
Then the generalized Fresnel equation (\ref{eq7x}) can be written as
\begin{equation}\label{Fresnel3D}
Mq_0^4+M^{\a\b}q_\a q_\b q_0^2+M^{\a\b\g\d}q_\a q_\b q_\g q_\d =0\,.
\end{equation}

For $f\ne 0$, this is a quadratic equation for $q_0^2$. Its solution
reads
\begin{equation}\label{fff}
q_0^2=\frac{-M^{\a\b}q_\a q_\b \pm\sqrt{\Delta}}{2M}\,,
\end{equation}
with the discriminant 
\begin{equation}\label{disc}
\Delta:= (M^{\a\b}q_\a q_\b)^2 -4MM^{\a\b\g\d}\,
q_\a q_\b q_\g q_\d\,.
\end{equation} 
This discriminant, upon substitution of (\ref{eq19xx}), vanishes.
Therefore we obtain a unique light cone
\begin{equation}\label{eq20}
fq_0^2-h(q_1^2+q_2^2+q_3^2)=0\,.
\end{equation}
The light velocity is positive, provided $f/h>0$. Thus, in the general
model as well as in the Preuss model, the light cone is single and the
effect of birefringence absent, see \cite{ROH}.  Certainly, this is a
result of the isotropy of the Friedmann model.  Accordingly, photons
would propagate isotropically but with a torsion dependent velocity
$v^2=f/h$. In astrophysical observations this effect could shows up in
a certain deviation from the cosmological redshift predictions of
general relativity.


\medskip

\noindent{\it Acknowledgment:\/} We are very grateful to Claus 
L\"am\-merzahl (Bremen) and to Yuri Obukhov (Cologne/ Moscow) for many
useful remarks.



\begin{thebibliography}{99}
\bibitem{conv1} Here $a,b,...=0,1,2,3$ are frame indices and $\vt^a$ represents the
coframe.
\bibitem{Reissner} See \cite{gron96}, Eq.(128). The Lagrangian is
  $\sim(1/2\ell^2)\, \left(T^a\wedge\right.$ $\left.\vartheta^b\right)
  \wedge{}^*\left(T_b\wedge\vartheta_a\right)+(1/2\kappa)\,
  R^{ab}\wedge{}^* R_{\a\b}+{}^*F\wedge F$, with $ R^{ab}$ as
  curvature, i.e., no explicit interaction term between gravity and
  electrodynamics was introduced. Still, the torsion turns out to be
  $T^{\hat r}=(Mr-2q^2)r^{-3}\; \vt^{\hat t}\wedge\vt^{\hat
    r}\,,\;{\rm etc.}$, with $t$ and $r$ as temporal and radial
  Schwarzschild coordinates, respectively. Obviously, the torsion does
  depend on the electric charge $q$.
\bibitem{conv} In \cite{ROH,gentle,book}, the twisted 2-form $H$ is
  decomposed according to $H=H_{ab}\vt^{ab}/2$.  Then the spacetime
  relation reads $\check{H}^{ab}=\chi^{abcd}F_{cd}/2$, with
  $\check{H}^{ab}=\epsilon^{abcd}H_{cd}/2$.  Correspondingly, up to a
  factor $\sqrt{-g}$, the constitutive tensors $\chi^{abcd}$ are the
  same.
\bibitem{Faraday} This should be carefully distinguished from the
  Faraday effect that has a different parity behavior; see, however,
  \cite{clausannalen}.
\bibitem{greek} We use Greek indices $\a,\b,...=1,2,3$ for spatial
  coordinates.
\end{thebibliography}
\end{document}